\documentstyle[prl,aps,multicol]{revtex}
\begin{document}
\draft 

\title{Shifts of Random Energy Levels by a Local Perturbation}

\author{I.L. Aleiner$^{1}$ and K.A. Matveev$^{2}$}

\address{$^{1}$NEC Research Institute,
  4 Independence Way, Princeton, NJ 08540\\
  $^{2}$ Duke University, Department of Physics, Durham, NC 27708-0305 }
\date{November 25, 1997} 
\maketitle

\begin{abstract}
  We consider the effect of a local perturbation on the energy levels
  of a system described by random matrix theory. An analytic
  expression for the joint distribution function of initial and final
  energy levels is obtained.  In the case of unitary ensemble we also
  find the two-point correlation function of initial and final
  densities of states.
\end{abstract}

\begin{multicols}{2}
  
The random matrix theory\cite{Wigner,Dyson1,Porter} of energy levels in
complex systems was developed in the fifties for the description of the
absorption spectra of large nuclei. In this approach one gives up any
attempt to study the position of each particular resonance, but instead
concentrates on the characteristics averaged over large number of
resonances. The positions of the resonances are identified with the
eigenvalues of some matrix, and the averaging is performed over the
elements of this matrix.

The statistical properties of the eigenvalues $\varepsilon_i$ of a
hermitian random matrix ${\hat H}$ of size $N\times N$ are completely
described by their joint distribution function $P(\{\varepsilon_i\})$,
which can be written in the following simplified form
\begin{equation}
P(\{\varepsilon_i\})\propto\prod_{i>j}
\left(\varepsilon_i - \varepsilon_j\right)^\beta.
\label{WD}
\end{equation}
Here the energies $\varepsilon_i$ are ordered: $\varepsilon_i <
\varepsilon_{i+1}$; the exponent $\beta =1,2,4$ for ensembles of
orthogonal, unitary and symplectic matrices, respectively.

Expression (\ref{WD}) vanishes when any two energy levels approach each
other; this effect is commonly referred to as {\it level repulsion}. This
repulsion can also be illustrated by rewriting $P(\{\varepsilon_i\})$ as
\begin{equation}
  \label{gas}
  P(\{\varepsilon_i\})=e^{-\beta E}, 
  \quad E=-\sum_{i>j} \ln(\varepsilon_i-\varepsilon_j).
\end{equation}
Thus $P(\{\varepsilon_i\})$ can be interpreted as Gibbs distribution
of a gas of classical particles at points $\varepsilon_i$ with logarithmic
repulsion between them. In physically interesting situations there must be
a finite average distance $\Delta$ between the particles (energy levels).
This is usually achieved by either introducing a parabolic confining
potential $\delta E=\alpha\sum_i \varepsilon_i^2$ or by confining the
particles to a circle\cite{Mehta}.

More recently the random matrix theory was applied to a number of
physically different systems, such as metallic grains
\cite{GorkovEliashberg,grains} and microwave cavities\cite{microwave}. In
these systems one can easily modify the matrix $\hat H$, e.g., by applying
magnetic field to the grain or by deforming the microwave cavity.  Such
modification can be described by a perturbation $\hat V$, and one is
usually interested in the correlations of the energy levels of the old and
new systems described by matrices $\hat H$ and $\hat H + \hat V$.

Dyson\cite{Brownian} suggested to describe such correlations in terms of
the viscous Brownian motion of the infinitely heavy particles (\ref{gas})
with the same logarithmic interactions between them. The external
perturbation in this approach plays the role of the fictitious time in
which the Brownian motion occurs. What remains then is to find the
distribution of the positions of all the particles after some time $t$,
provided initial distribution (\ref{WD}). Time $t$ can be related to the
characteristic value of the potential $\hat V$, so that the parametric
correlations are universal functions of only one parameter.

However the Brownian-motion approach is not always applicable.  Consider a
perturbation of the general form
\begin{equation}
\hat{V}= N\sum_{i=1}^{N} v_i|i\rangle\langle i|,
\label{perturbation}
\end{equation}
where $|i\rangle$ form a complete set of states, and the matrix
dimension $N$ is included for proper definition of limit $N\to
\infty$.  For the Brownian-motion model to be
applicable\cite{Brownian,Chalker}, the condition $v_i \ll {\Delta}$
must hold, where ${\Delta}$ is the mean level spacing. Since $N\gg1$,
the sum $\sum_iv_i^2$ which has the meaning of the fictitious time $t$
in the Brownian-motion picture, can still be arbitrarily large.

In a number of interesting physical situations one deals with a local
perturbation described by Eq.~(\ref{perturbation}) with
$v_i=v\delta_{i1}$, where $v$ is not necessarily small. An example of such
perturbation is a short-range impurity in a metallic grain. When such an
impurity is added to the system, its levels $\varepsilon_i$ shift to new
positions $\lambda_i$. The new many-particle ground state $|\Psi\rangle$
has a rather small overlap with the old one, $|\Phi\rangle$; this
phenomenon\cite{Anderson} is called orthogonality catastrophe. The overlap
can be expressed in terms of the old and new energy levels:
\begin{equation}
\left|\langle\Psi |\Phi\rangle\right|^2=
\prod_{i=1}^M\, \prod_{j=M+1}^N
\frac{\left(\lambda_j-\varepsilon_i\right)
\left(\varepsilon_j-\lambda_i\right)}
{\left(\lambda_j-\lambda_i\right)
\left(\varepsilon_j-\varepsilon_i\right) },
\label{orthogonality}
\end{equation}
where $M$ is the number of electrons in the system\cite{remark}. Therefore
to treat the orthogonality catastrophe in a metallic grain one needs the
knowledge of the joint distribution function $P(\{\varepsilon_i\},
\{\lambda_i\})$ of {\it both\/} old and new energy levels. 

The determination of this joint distribution function is the main subject
of this paper. The Brownian-motion model is not applicable in this case,
but a closed analytic expression for $P(\{\varepsilon_i\},
\{\lambda_i\})$ can be found directly. For the orthogonal, $\beta=1$, and
unitary, $\beta=2$, ensembles we will show that
\begin{eqnarray}
P(\{\varepsilon_i\}, \{\lambda_i\})&\propto&
\frac{{\prod_{i>j}} (\varepsilon_i - \varepsilon_j) 
                                 (\lambda_i - \lambda_j)} 
     {{\prod_{i,j}} 
       \left|\varepsilon_i - \lambda_j\right|^{1-\beta/2}}
\nonumber\\ 
&& \times \textstyle
\exp\left[-\frac{\beta}{2v}
                   \sum_i(\lambda_i-\varepsilon_i)\right].
\label{result}
\end{eqnarray}
Energy levels in Eq.~(\ref{result}) are constrained by the condition
\begin{equation}
  \begin{array}{cl}
 \varepsilon_i \leq \lambda_i\leq \varepsilon_{i+1},
  \quad \mbox{if $v>0$},\\[1ex]
 \varepsilon_{i-1} \leq \lambda_i\leq \varepsilon_{i}, 
  \quad \mbox{if $v<0$}.
  \end{array}
 \label{constraint}
\end{equation}
Equation (\ref{result}) is the central result of this paper.

To derive Eq.~(\ref{result}) we need to relate the eigenvalues
$\lambda_j$ of the perturbed matrix $\hat{H}+ vN|v\rangle\langle v|$ to
the unperturbed eigenvalues $\varepsilon_i$ and eigenfunctions
$|i\rangle$; here $|v\rangle$ is an arbitrary vector. This is easily
accomplished:
\begin{equation}
  \sum_i \frac{A_i}{\lambda_j-\varepsilon_i}=\frac{1}{vN},
  \quad A_i\equiv \left|\langle v|i\rangle\right|^2.
 \label{newold}
\end{equation}
Equation (\ref{newold}) enables one to find $P(\{\varepsilon_i\},
\{\lambda_i\})$ given the joint distribution function of unperturbed
eigenvalues and eigenfunctions. Since the distributions of eigenvalues and
eigenvectors in the random matrix theory are uncorrelated, we have
\begin{equation}
  \label{algebra}
  P(\{\varepsilon_i\}, \{\lambda_i\}) = 
      P(\{\varepsilon_i\}) p(\{A_i\}) 
      \left|\det\left[
        \frac{\partial A_i}{\partial \lambda_j}\right]\right|.
\end{equation}
Here $P(\{\varepsilon_i\})$ is the distribution function (\ref{WD}) of
the unperturbed energy levels, and $p(\{A_k\})$ is the eigenvector
distribution function, which at $N\to\infty$ is given by
Porter-Thomas\cite{PorterThomas} formula:
\begin{equation}
  p(\{A_i\})=
    \prod_i \frac{N}{(2\pi NA_i)^{1-\beta/2}}
    \exp\left(-\frac{\beta}{2}NA_i\right),
\label{p}
\end{equation}
where $\beta=1$ or 2.  Finally, the last factor in Eq.~(\ref{algebra}) is
the Jacobian of the transformation from the eigenvector variables $A_i$ to
the new energies $\lambda_j$.

To find the distribution function (\ref{algebra}) we do not need to solve
the equation (\ref{newold}) with respect to $\lambda_j$. In order to find
the Jacobian in Eq.~(\ref{algebra}) one only has to solve (\ref{newold})
with respect to $A_i$. The latter is a much simpler problem since equation
(\ref{newold}) is linear in $A_i$, and the solution can be expressed in
terms of Cauchy determinants. This readily yields
\begin{equation}
A_i=\frac{1}{vN}
\frac{\prod_{j=1}^N
\left(\lambda_j-\varepsilon_i\right)}
{\prod_{j\neq i}
\left(\varepsilon_j-\varepsilon_i\right)}.
\label{Ak}
\end{equation}
By definition all $A_i$ are positive, see Eq.~(\ref{newold}). This
immediately gives constraint (\ref{constraint}).  It follows from
Eq.~(\ref{Ak}) that 
\begin{equation}
  \label{identity}
  \frac{\partial A_i}{\partial \lambda_j} 
    = \frac{A_i}{\lambda_j - \varepsilon_i}.
\end{equation}
As a result the Jacobian in Eq.~(\ref{algebra}) is reduced to a Cauchy
determinant, and we obtain
\begin{equation}
  \det\left[\frac{\partial A_i}{\partial \lambda_j}\right] 
  = \frac{1}{\left(Nv \right)^N}
    \frac{\prod_{j>i} \left(\lambda_j-\lambda_i\right)}
    {\prod_{j>i} \left(\varepsilon_j-\varepsilon_i\right)}.
\label{Jresult}
\end{equation}

In order to express the Porter-Thomas distribution function (\ref{p}) in
terms of the energies $\varepsilon_i$ and $\lambda_j$ we need to evaluate
the sum $\sum_i A_i$. To this end we sum up both sides of identity
(\ref{identity}) over $i$, and using Eq.~(\ref{newold}), find
\[
 \frac{\partial}{\partial{\lambda_j}}\sum_i A_i=\frac{1}{Nv},
  \quad j=1,\dots,N. 
\]
We therefore conclude that the sum of $A_i$ is a linear function of all
$\lambda_i$. The constant can be determined by noticing that according to
Eq.~(\ref{Ak}) at $\lambda_i\to\varepsilon_i$ we have $A_i\to0$. Thus
\begin{equation}
\sum_i A_i=\frac{1}{Nv}\sum_i (\lambda_i - \varepsilon_i). 
\label{sum}
\end{equation}

Finally, we substitute Eqs.~(\ref{WD}), (\ref{p}), and (\ref{Jresult})
into Eq.~(\ref{algebra}), and with the help of Eqs.~(\ref{Ak}) and
(\ref{sum}) get the joint distribution function Eq.~(\ref{result}).
Strictly speaking the result (\ref{result}) is valid only in the limit
$N \to\infty$. This is the physically most interesting regime where
the properties of the system are universal.

If one is interested in non-universal corrections associated with the
finite size of the matrix, the Porter-Thomas distribution (\ref{p})
should be replaced by\cite{Mehta}
\begin{equation}
p\left(\left\{A_i\right\}\right)= \frac{\Gamma \left(\beta 
N/2\right)}{\Gamma\left(\beta/2\right)^{N}}
{\left(\prod_i A_i\right)^{\beta/2-1}}
\delta\left[1-\textstyle{\sum_i}A_i\right].
\label{pe}
\end{equation}
As a result, the exponential factor in Eq.~(\ref{result}) is replaced
by $\delta[1-(Nv)^{-1}\sum_i(\lambda_i-\varepsilon_i)]$. However, for
practical calculations the distribution function in the from
(\ref{result}) is more convenient.

Our result (\ref{result}) contains complete information about
distribution of the old and new energy levels of the system. In
applications, such as analysis of experimental spectra, one often
needs only a small part of this information, which is contained in
$n$-level correlation functions. The most important of them is the
two-point correlation function of the old and new densities of states
$K_2$. We define this quantity as
\[
K_2(s)=\frac{1}{\rho^{2}(E)}
\sum_{ik}\left\langle\delta\left(E+\frac{s}{2}-\lambda_i\right)
\delta\left(E-\frac{s}{2}-\varepsilon_k\right)
\right\rangle.
\]
$K_2(s)$  has the meaning of the probability to find a new level at a
distance $s$ from a given old level.  The correlation function $K_2$
does not depend on energy $E$, provided that $s$ is much smaller than
the characteristic energy scale over which average density of states
$\rho(E)= \sum_i\langle\delta\left(E-\lambda_i\right)\rangle$ varies.

We have been able to obtain a compact analytic expression for $K_2(s)$ for
the unitary ensemble only, and we outline the derivation below. Because
$K_2$ does not depend on the particular shape of $\rho(E)$, it is
convenient to get rid of energy dependence of $\rho (E)$ by adopting the
circular ensemble of Dyson\cite{Dyson1}, where all $N$ levels are put on
the circle of unit radius.  Equation (\ref{result}) for $\beta = 2$ then
takes the form
\begin{equation}
P_c=\left[\prod_{i>j} 
4\sin\left(\textstyle{\frac{\varepsilon_i - \varepsilon_j}{2}}\right)
\sin \left(\textstyle{\frac{\lambda_i - \lambda_j}{2}}\right)\right]
e^{-\frac{1}{v}\sum_i(\lambda_i-\varepsilon_i)}. 
\label{circular}
\end{equation}
Here the energies are measured in dimensionless units and defined
within the interval $\left[-\pi,\pi\right]$; we have also omitted the
normalization constant.  Mean level spacing in such model is given by
${\Delta}=2\pi/N$.

\begin{mathletters}
We express function $K_2$ in terms of the functional derivative
\begin{equation}
K_2= \left.\frac{\Delta^2}
                {I\left[{\cal A},{\cal B}\right]}
     \frac{\delta^2I\left[{\cal A},{\cal
                B}\right]}
       {\delta{\cal A} \delta{\cal B}}
     \right|_{{\cal A},{\cal B}=0}
\label{derivative}
\end{equation}
of the generating functional for $v>0$, (negative $v$ are considered
analogously)
\begin{eqnarray}
I&=&\int_{-\pi}^{\pi}\!\!\!
d\varepsilon_1\! \int_{\varepsilon_1 }^\pi\!\!\! d\lambda_1 
\!\int^{\pi}_{\lambda_1}\!\!\!
d\varepsilon_2\! \int_{\varepsilon_2 }^\pi\!\!\! 
d \lambda_2\dots\!
\int_{\lambda_N }^\pi 
\!\!\! d \lambda_N
P_c
\nonumber\\
 &&\times\prod_k\left(1+{\cal A}(\varepsilon_k)\right)
\left(1+{\cal B}(\lambda_k)\right)
\label{functional}
\end{eqnarray}
\end{mathletters}

Following the procedure similar to that of Ref.~\cite{Dyson1}, we find
that the generating functional $I$ can be rewritten as determinant of
certain matrix $I=\det\, \hat{F}$, where
\begin{equation}
F_{kl}\!=\!\int_{-\pi}^\pi\!\!\! d\varepsilon\!
\int^\pi_\varepsilon\!\!\!
 d\lambda\ 
e^{(v^{-1}+ik)\varepsilon -(v^{-1}+il)\lambda}
\left[1+{\cal A}(\varepsilon)\right]
\left[1+{\cal B}(\lambda)\right]
\label{matrix}
\end{equation}
We then expand the determinant up to the second order in small sources
${\cal A}$ and ${\cal B}$. This expansion requires the knowledge of
the matrix $\hat F^{-1}$ at ${\cal A}={\cal B}=0$, which in the
limit $N\to\infty$ takes the form
$F^{-1}_{kl}=\delta_{kl}\frac{v^{-1}+ik}{2\pi}$. Substituting the
result in Eq.~(\ref{derivative}) we obtain after simple
algebra\cite{Simons}
\begin{equation}
K_2=1 -
\left[
\theta(r) 
-\!\!
\int_{-\infty}^{r}\!\!\! dr^\prime
e^{\frac{r^\prime\Delta}{v}}
\frac{\sin\pi r^\prime}{\pi r^\prime}
\right]
\frac{\partial}{\partial r}
\left[
e^{-\frac{r\Delta}{v}}
\frac{\sin\pi r}{\pi r}
\right],
\label{R2result}
\end{equation}
where $r=s/\Delta$ is the energy in the units of the level spacing.
The result (\ref{R2result}) is valid for positive $v$; for $v<0$ 
one should substitute $r \to -r$ in Eq.~(\ref{R2result}). 

Let us now discuss asymptotic behavior of the two-point correlation
function (\ref{R2result}). In the limit of vanishing perturbation
$v\to 0$, we immediately obtain $K_2=\delta(r)+1-(\sin\pi r/\pi r)^2$
which is the well-known result for the two-point correlation function
of the unitary ensemble\cite{Porter,Mehta}.  In the limit $r \gg 1,
v/\Delta$, the integration in Eq.~(\ref{R2result}) can be easily
performed and we find
\begin{equation}
K_2 = 1- \frac{\sin^2\left(\pi r - \delta \right)}{(\pi r)^2},
\label{larger}
\end{equation}
where $\delta = \arctan \left(\pi v/\Delta\right)$ is the phase shift
of the scattering off of the impurity. The sequence of periodic maxima
of the correlation function (\ref{larger}) is a signature of the level
repulsion, and the average shift $\delta$ of the new levels with
respect to the old ones is consistent with the Friedel sum rule
$\langle \lambda_i - \varepsilon_i\rangle/\Delta = {\delta}/{\pi}$.

In conclusion, we studied the statistics of shifts of eigenvalues of a
random Hamiltonian by a local perturbation of arbitrary strength. Despite
the fact that the conventional Brownian motion model is not applicable, we
have found the whole joint distribution function of old and new levels,
Eq.~(\ref{result}), and the correlator of the old and new densities of
states (\ref{R2result}). Using Eq.~(\ref{orthogonality}), these results
can be applied to the study of orthogonality catastrophe in small metallic
grains.

\end{multicols} 
\end{document}